\documentstyle[twoside,fleqn,espcrc2,epsf]{article}
\epsfverbosetrue
 
\newcommand{\AmS}{{\protect\the\textfont2
  A\kern-.1667em\lower.5ex\hbox{M}\kern-.125emS}}
 
\title{QED on a momentum lattice
\thanks{Work supported in part by the National Science Fundation under grant
NSF-PHY92-00148\hfill\break
\dag Present address: Dept. of Physics and Astronomy,
 University of Kentucky, Lexington, KY 40506 
}}
 
\author{J.B. Kogut and J.-F. Laga\"e \footnote{}\\
\vspace{0.25cm}
Department of Physics, University of Illinois,
1110 West Green Street, Urbana IL 61801, USA}

\begin{document}
 
\begin{abstract}
We investigate the possibility of doing momentum space lattice simulations
as an alternative to the conventional method. The procedure is introduced
and tested for quenched QED2 and quenched QED3. Interesting physical 
applications to unquenched QED3 and quenched QED4 are also briefly discussed.
\end{abstract}
 

\maketitle

\section{Introduction}
The basic idea of the momentum space lattice method is very straigthforward:
one just studies continuum QED on the torus (or more generally on any compact
manifold). Fourier expansion then leads to a discretized problem and the number 
of degrees of freedom can be made finite by imposing an ultraviolet cut-off.
For simplicity, we will consider here that the length $L$ of the torus (size
of the box) is the same in each direction. However, finite temperature type
situations can also easily be handled.

After gauge fixing (to the Feynman gauge) and Fourier expansion, the Euclidian
action of QED in d dimensions can be recast in the form:
\begin{eqnarray*}
S &=& -{1 \over 2}\beta\sum_q \theta_{\mu}^*(q)(2\pi q)^2 \theta_{\mu}(q)\\
  &+& \sum_k \overline{\chi}(k)[ \gamma_{\mu}(2 \pi k_{\mu}) + \rho ] \chi (k)
\\
  &-& \sum_{k,k^\prime} \overline{\chi}(k) \gamma_{\mu} \theta_{\mu}(k-k^\prime)
      \chi (k^\prime)
\end{eqnarray*}
where, instead of the continuum fields $\psi$ and $A_\mu$, we have used the
following dimensionless variables:
$$ \chi(k)=L^{(d+1) \over 2} \psi(k) \hskip .5cm
   \theta_{\mu}(q) = eL^{(d-1)}A_{\mu}(q)$$
and $ \beta=L^{(d-4)} / {e^2} \hskip .5cm \rho=mL $.
In the above formula, the $q_\mu{}'s$ take integer values and the $k_\mu{}'s$
are integers or half-integers depending on whether the boundary conditions
for the fermions are periodic or antiperiodic. On an $N^d$ lattice ($\equiv$
definition) the sums over $k$ and $k^\prime$ are truncated in such a way that
N positive frequency modes are included in each direction. $q$ then takes
all the possible values of the momentum transfer between $k$ and $k^\prime$.

The method has two main advantages: 1) the coupling of the fermions to the
gauge fields is non-compact and 2) there is no fermion doubling. The first
is crucial for physical applications to QED3 and QED4 as will be discussed
below. The second is simply a consequence of the fact that what we consider
is nothing more than a particular formulation of continuum QED.
The method also has two potential problems, which we should discuss here.
They are: 1) The fact that the fermionic matrix is no longer sparse in
momentun space (due to the convolution with the gauge field) and 2)
The breaking of gauge invariance induced by imposing an ultraviolet cut-off.
The first is easily dealt with by using repeated Fast Fourier Transform in
matrix multiplications. This brings the computer time requirements from
$O(N^2)$ to $O(N\ln N)$, which is still $\sim\ln N$ slower than the position
space method; however, because of the non-compact character of the coupling
mentioned above, smaller lattices are generally sufficient in momentum space,
so that on the overall, we can still gain. The problem of gauge invariance
is well known from conventional perturbation theory in quantum electrodynamics:
$\Pi_{\mu\nu}$ is not transverse when computed with an ultraviolet cut-off.
The cure to this problem is to use a gauge invariant regularization: for
example, Pauli-Villars. In the case of QED2 and QED3, the procedure simply
consists in dividing the usual fermionic determinant by the determinant of
a heavy fermion of mass $M$ and taking the limit $\Lambda\rightarrow\infty$
(number of modes going to infinity) before letting $M\rightarrow\infty$.
In 4 dimensions, the procedure is slightly more involved and would include
two regularizing determinants \cite{ZUBER}. Coming back to the previous case,
the implementation of the algorithm is straightforward: in the Hybrid Monte
Carlo algorithm, one replaces the usual pseudo-fermion potential by 
$\phi^*A^\dagger_M ( A_m^{\vphantom{\dagger}} A^\dagger_m)^{-1}
A_M^{\vphantom{\dagger}} \phi$.
As far as computer time is concerned, the additional multiplication by $A_M$
doesn't make much difference.

\section{QED2}

\begin{figure}[htb]
\vspace{50pt}
\epsfbox[0 0 213 100]{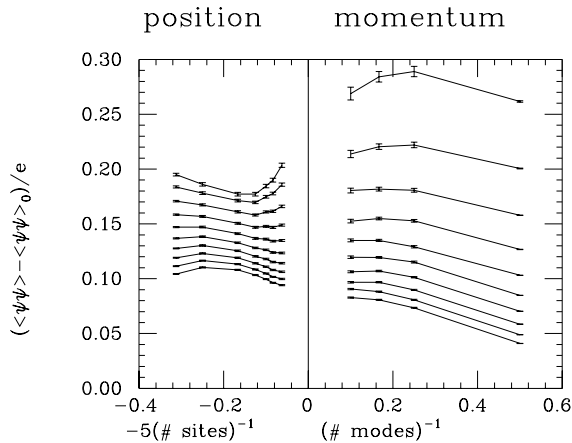}
\caption{A comparison between position and momentum space measurements of
the chiral condensate. $eL=20$ and $m/e$ ranges from 0.02 to 0.2 (from top
to bottom)}
\end{figure}

Because it is computationally cheap, QED2 has been considered as a test case.
We restricted our attention to the zero topological sector of the theory, i.e.
configurations for which the flux of the electric field through the torus is
zero (whereas in general $\phi=2\pi n$). This restriction would be a drastic
step in the full Schwinger model (where the chiral condensate comes from
non-trivial topology \cite{WIPF}). However, in the quenched case to be discussed
here, the theory resulting from this approximation remains interesting. We
measured both $\langle\overline{\psi}\psi\rangle$ and the mass gap on lattices
of various ``physical size'' (eL=20,30,40) for various number of modes
($2\times2,4\times4,6\times6,10\times10$) and for various values of the bare
fermion mass (ranging from ${m / e} = 0.02$ to ${m / e} = 0.2$). Some
of our results for $\langle\overline{\psi}\psi\rangle$ are compared on
fig. 1 with those obtained from the position space non-compact formulation
of  QED2. Although in both cases the extrapolation to an infinite ultraviolet
cut-off appears rather difficult, an agreement between the two methods seems
possible. It would certainly be interesting to investigate this in greater
detail particularly in relation with the claim \cite{DIV} that the chiral
condensate should diverge in the quenched Schwinger model (remember that we
have only the zero topological sector here). We have also measured the mass
gap in this model. The results are presented in Fig. 2 in the form of
$M(p)/|p|$ where $M$ is the dynamical fermion mass and p is the lowest
momentum available on the lattice ($p_0=\pi/L,p_1=0$). As one should
expect, the results converge faster with increasing number of modes on the
smaller lattices. However, the signal for chiral symmetry breaking becomes
better and better as the physical size of the lattice is increased.
Extremely good signals for chiral symmetry breaking, are in fact a general
feature of the momentum space lattice method, as will be verified below.

\begin{figure}[t]
\vspace{30pt}
\epsfbox[0 0 213 100]{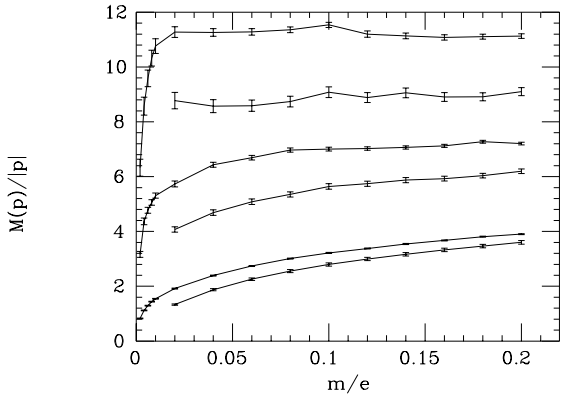}
\caption{Dynamical versus bare fermion mass on $2\times2$ and $6\times6$
lattices of size eL=40(top), 30(middle) and  20(bottom)}
\end{figure}

\section{QED3}

As is well known, the breaking of chiral symmetry in unquenched or finite
temperature QED3 relies entirely on weak long range interactions. As a
consequence, conventional lattice studies of this problem have been plagued
by strong finite size effects \cite{LAT92} and quantitative measurements of the
critical temperature or a determination of the phase diagram as a function of
the number of flavors have remained beyond reach. The difficulty essentially
comes from the fact that since large lattices are needed and the number of
sites is limited by computer requirements, the lattice spacing $a$ is
necessarily large and one falls inevitably into the strong coupling regime
(remember that $\beta^{-1}=e^2a$ in (2+1) dimensions). The interest of
applying the momentum space lattice method to this problem, comes from the fact
that, in this formulation, the coupling of the fermions to the gauge field is
non-compact. Therefore, one can increase the ``physical size'' of the lattice
while remaining relatively close to the continuum limit. Our preliminary
results so far mainly concern the quenched case where we have checked that
the method gives a very clear signal for chiral symmetry breaking :
In the relevant range, the dynamical mass at zero momentum is always much
larger than the bare fermion mass.

\section{QED4}
One of the most striking features of the numerical simulations of quenched
non-compact QED4 is that the results do not agree with the analytical
prediction of Miransky. It has been argued that this is a consequence
of induced 4-fermi interactions on the lattice and shown that the critical
exponent are expected to vary continously along the critical line in the
$(\alpha,G)$ plane \cite{HANDS}. However, until now, no procedure has been found
for modifying the lattice action in such a way that the results of the
simulation would move (along the critical line) in the direction of the
Miransky-Bardeen-Leung-Love point. We would like to argue here that the
momentum space lattice method is a good candidate for performing this task.
The first step consists in identifying the origin of the 4-fermi interactions
on conventional lattices. An obvious candidate is revealed by performing
exactly the integration over the gauge field in the path integral (which
is straightforward for the non-compact action after gauge fixing). The
procedure gives rise to an effective action which contains a 4-fermi
interaction, a ``tadpole improved'' kinetic energy term, the usual
current-current interaction and higher order interactions. The derivation
shows clearly that the 4-fermi term is a direct consequence of the compact
character of the coupling between fermions and gauge fields and would therefore
not be there on a momentum lattice. As a consequence, we expect that the
momentum space method will give results which are much closer to those of
Miransky or at least will allow us to check whether these results survive
the introduction of vertex corrections.

\section{Conclusion}
We have seen that the procedure of momentum space lattice simulation is
practical and gives a very clear signal for chiral symmetry breaking for
quenched QED both in 2 and 3 dimensions. We hope to be able to report soon
on physical applications to quenched QED4. The unquenched theory in lower
dimensions is also currently being tested.

\end{document}